\documentclass[english]{elsarticle}
\usepackage{lmodern}

\usepackage[T1]{fontenc}
\usepackage[latin9]{inputenc}
\usepackage{geometry}
\usepackage{units}
\usepackage{amstext}
\usepackage{graphicx}
\usepackage{esint}

\makeatletter

\providecommand{\tabularnewline}{\\}

\journal{Journal of Alloys and Compounds}

\makeatother

\usepackage{babel}
\begin{document}

\title{Synthesis, single crystal growth and properties of Sr$_{5}$Pb$_{3}$ZnO$_{12}$}

\author[pg]{M.J.~Winiarski\corref{cor1}}

\ead{mwiniarski@mif.pg.gda.pl}

\author[pg]{N. A. Szreder}

\author[pg]{R. J. Barczy\'{n}ski}

\author[pg]{T. Klimczuk\corref{cor1}}

\ead{tomek@mif.pg.gda.pl}

\cortext[cor1]{Corresponding authors: }

\address[pg]{Faculty of Applied Physics and Mathematics, Gda\'{n}sk University
of Technology, ul. Narutowicza 11/12, 80-233 Gda\'{n}sk, Poland}
\begin{abstract}
The novel Sr$_{5}$Pb$_{3}$ZnO$_{12}$ oxide was synthesized by the
solid-state reaction method. The crystal structure was studied by
means of the powder x-ray diffraction Rietveld method and was found
to be similar to 3 other previously known Sr$_{5}$Pb$_{3}$MO$_{12}$
compounds (M = Co, Ni, Cu). Crystals of several hundred microns in
size of the new phase were grown in molten sodium chloride and imaged
using confocal optical and scanning electron microscopy. Electrical
properties were studied using the impedance spectroscopy technique.
It was found that Sr$_{5}$Pb$_{3}$ZnO$_{12}$ is a dielectric material
with rather high relative permittivity ($\varepsilon_{r}=22$ at 300K)
and with activation energy of the dielectric relaxation process $E_{A}=0.80(4)\: eV$.
The heat capacity studies reveal the Debye temperature $\Theta_{D}=324(1)\: K$.\end{abstract}
\maketitle

\section{Introduction}

Oxide compounds that contain chains of transition metals are expected
to reveal interesting electronic and magnetic properties\citep{nguyen_design_1996}.
Indeed one-dimensional Heisenberg antiferromagnetism was observed
in Ca$_{4}$Cu$_{5}$O$_{10}$\citep{imada_metal-insulator_1998},
Peierls instability was found in CuGeO$_{3}$\citep{hase_observation_1993}
and spinon-holon separation in SrCuO$_{2}$\citep{kim_observation_1996}.
In the quasi-one-dimensional BaVS$_{3}$ sulfide compound, a metal-semiconductor
transition and Luttinger behavior were reported \citep{nakamura_metal-semiconductor_1994}.

One of the oxide systems with infinite chains of metals is Sr$_{5}$Pb$_{3}$MO$_{12}$
(M = Co, Ni, Cu). The first discovered Sr$_{5}$Pb$_{3}$MO$_{12}$
(531-M) compound was Sr$_{5}$Pb$_{3}$CuO$_{12+\delta}$ found as
an impurity phase in the high-temperature superconductor BSCCO\citep{kim_new_1990}.
It has been synthesized by Babu, et al. \citep{babu_neutron_1991}
and magnetic properties were studied by Yamaura, et al. \citep{yamaura_crystal_2001}.
To date, two more 531-M compound were synthesized with M = Co, Ni,
all having similar structure to 531-Cu\citep{yamaura_synthesis_2002,prior_pentastrontium_2004}.
Both 531-Cu and 531-Co are known to be dielectrics but no extensive
studies of electrical properties were conducted. In Sr$_{5}$Pb$_{3-x}$Bi$_{x}$MO$_{12}$
, Bi substitution for Pb does not significantly change the electrical
conductivity but does change the one-dimensional antiferromagnetic
properties. \citep{yamaura_crystal_2001}

In this study, polycrystalline samples of a novel 531-Zn compound
were synthesized by solid-state reaction and electrically characterized.
The structure of 531-Zn was studied using powder x-ray diffraction
via the Rietveld refinement method.

\section{Materials and methods}

The polycrystalline Sr$_{5}$Pb$_{3}$ZnO$_{12}$ (531-Zn) sample
was prepared by high-temperature solid state reaction. The precursor
compounds: SrCO$_{3}$ (99.9\%, Chempur), Pb$_{3}$O$_{4}$ (99\%,
Aldrich) and ZnO (99.9\%, Aldrich) were weighted in stoichiometric
amounts (assuming 5:3:1 Sr:Pb:Zn ratio), well ground and mixed using
agate mortar and pestle and pressed into pellets, approx. 2 cm diameter,
using a hydraulic press (pressure applied did not exceed 50 MPa) to
obtain about 0.8 g of the 531-Zn compound. Pellets were placed in
dense alumina crucible and annealed in a furnace at 750$\textdegree$C
for 36 hours in air. The first step of the synthesis was intended
to decompose the strontium carbonate without completing the solid-state
reaction. The x-ray diffraction pattern measured on the powdered samples
showed the presence of 531-Zn as well as SrPbO$_{3}$ phase \citep{keller_zur_1975,grazulis_crystallography_2009}
(about 30\%). Pellets were then reground, repressed into smaller pellets
(approx. 0.5 cm diameter, pressures applied to the sample did not
exceed 300 MPa), placed in the same crucible used before and annealed
in a furnace at 800$\textdegree$C for 36 hours in air. After this
step, pellets were ground and preliminary x-ray diffraction measurement
revealed that the sample was nearly single-phase, with minor quantities
of SrPbO$_{3}$ impurity possibly present. The powder was then repressed
into pellets and annealed at 850$\textdegree$C for 36 hours in air,
and the product was found to be single phase 531-Zn. The same methodology
was then successfully employed to synthesize polycrystalline pellets
of 531-Cu and 531-Co for impedance spectroscopy measurements. CuO
(99.5\%, Reachim) and Co$_{3}$O$_{4}$ (99.7\%, Alfa Aesar) precursor
materials were used, respectively. The other reactants used were the
same as for 531-Zn.

To determine details of the crystal structure of the synthesized 531-Zn
phase, a specimen was studied by powder x-ray diffraction in Bruker
D8 FOCUS diffractometer with a Cu K$\alpha$ radiation source. The
diffraction pattern was analyzed by the Rietveld profile refinement
method\citep{rietveld_profile_1969} using the FullProf 5.3 program\citep{carvajal_fullprof:_1990}.
The background was modelled using linear interpolation between 74
points with refinable intensity values.

To measure the oxygen content of the material, iodometric titration
was conducted on a sample prepared as described previously, divided
into 7 portions of approx. 0.2 g each\citep{harris_preparation_1987,scheurell_determination_1991}.
Samples were dissolved in 100 ml of 0.33M hydrochloric acid with approx.
7 g of potassium iodide and then titrated with 0.1 M sodium thiosulphate
using starch as indicator (starch was added to the solution close
to the endpoint of titration to avoid precipitation of iodine-starch
complex\citep{karppinen_oxygen_2002}). 

Iodine liberation reaction proceeds as follows:

\begin{equation}
Pb^{4+}+2I^{-}\rightarrow Pb^{2+}+I_{2}
\end{equation}

\begin{equation}
Pb^{2+}+2I^{-}\rightarrow PbI_{2}\downarrow
\end{equation}

The second step of reaction resulted in formation of yellow precipitate
of a similar appearance to 531-Zn powder. To determine if dissolution
of 531-Zn was complete, XRD measurements were done on the filtered
residue, which showed that the precipitate did not consist of 531-Zn,
with only lead(II) iodide (PbI$_{2}$) present.

\begin{figure}
\begin{centering}
\includegraphics[width=10cm]{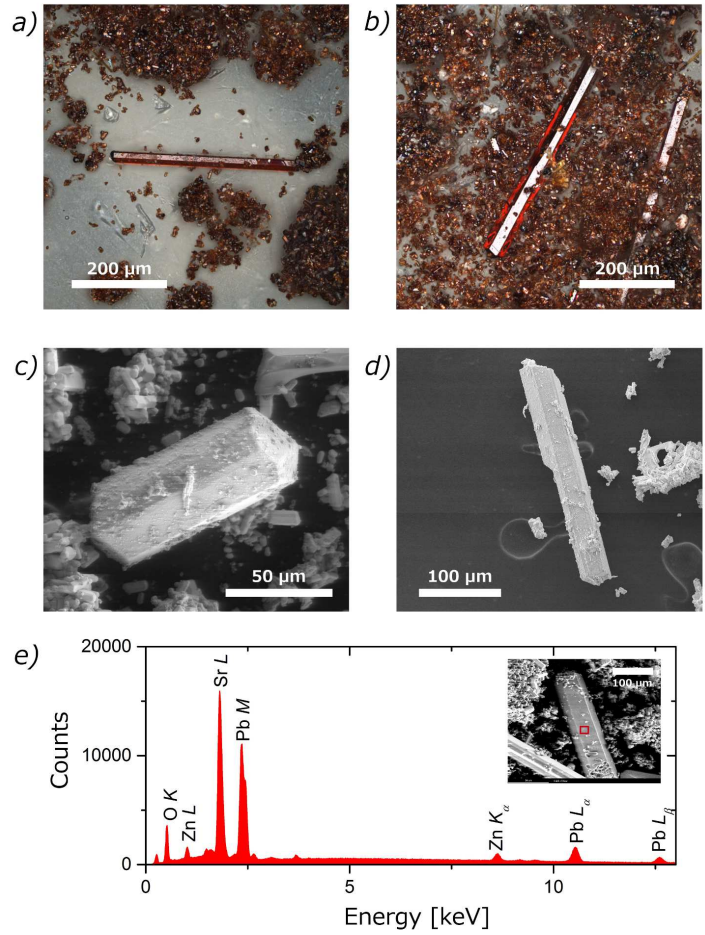}
\par\end{centering}

\centering{}\caption{CSLM (a, b) and SEM (c, d) micrographs of 531-Zn single crystals with
sample EDS spectrum for the single crystal (e). The inset of (e) shows
the single crystal from which the spectrum was collected with the
scanned area indicated by a red rectangle.}
\label{fig:PhotographsSEM_CSLM}
\end{figure}

The crystal growth mineralization method was adapted from the study
on 531-Ni\citep{prior_pentastrontium_2004}. Approximately 0.2 g of
531-Zn polycrystalline powder obtained after the second annealing
(800$\textdegree$C) was ground and mixed together with approx. 2
g of sodium chloride employed as a mineralizer. The powder was put
in an alumina crucible covered with a lid within another crucible
turned upside-down. The crucible was placed in an alumina boat and
put into a chamber furnace. The temperature program used was: 1) ramping
to 950$\textdegree$C at 180 C/h, 2) holding 950$\textdegree$C for
3 hours to homogenize the sample, 3) Slow cooling at 6$\textdegree$C/h
to 790$\textdegree$C (below the NaCl melting point) and switching
the furnace off. The contents of the crucible were then rinsed with
distilled water, filtered and dried. Then it was examined using Scanning
Electron Microscope (FEI Quanta) working in low-vacuum environmental
mode (chamber pressure 130 Pa, water vapor) and using Olympus LEXT
OLS4000 confocal scanning laser microscope (CSLM). The red-brown powder
obtained forms small needle-like crystals that have a rod-type shape
with clearly visible hexagonal faces (Fig. \ref{fig:PhotographsSEM_CSLM}).
The largest of the single-crystals were approximately 700 microns
long and 70 microns wide. 

EDS spectroscopy was conducted in low-vacuum conditions due to heavy
specimen charging, with the electron energy set to 20 keV. EDS spectra
were collected on 19 points for 9 distinct single crystals with an
acquisition time of 100 s. Further analysis was performed by using
EDAX TEAM$^{TM}$ software by means of a standardless analysis with
the \textit{eZAF} quantization method. Fig. \ref{fig:PhotographsSEM_CSLM}
(e) shows a sample EDS spectra collected for a single crystal.

Although the crystal growth was successful, the crystals were too
small to measure the physical properties. The crystal growth process
parameters (especially the cooling rate) should be optimized to obtain
crystals of larger dimensions, suitable for physical property measurements.

For the dielectric studies, a small cylindrical pellet (diameter 4.78
mm, height 0.66 mm) of 531-Zn was polished and gold contacts were
sputtered on both its basal surfaces. A mask with a hole with slightly
smaller diameter was used to protect the lateral surface of the sample.
A sample prepared this way was used for impedance spectroscopy measurements
conducted with a Novocontrol Concept 40 broadband dielectric spectrometer
in a temperature range from $-120$ to $+150\textdegree C$ (step
$30\textdegree C$) and frequency range 10 mHz-10 MHz. The applied
voltage (RMS) used was set to 1.0 V. Nyquist plots resulting from
the measurements were fitted with equivalent circuit using Scribner
ZView 2 program. The parameters of the circuit elements were refined
using the least squares method.

Heat capacity measurements were performed in a Quantum Design Physical
Property Measurement System using the standard relaxation calorimetry
method. A small piece (about $6\: mg$) of sample was measured in
the temperature range of $1.9$ to $300\: K$.

\section{Results and discussion}

\begin{figure}
\begin{centering}
\includegraphics[width=12.1cm]{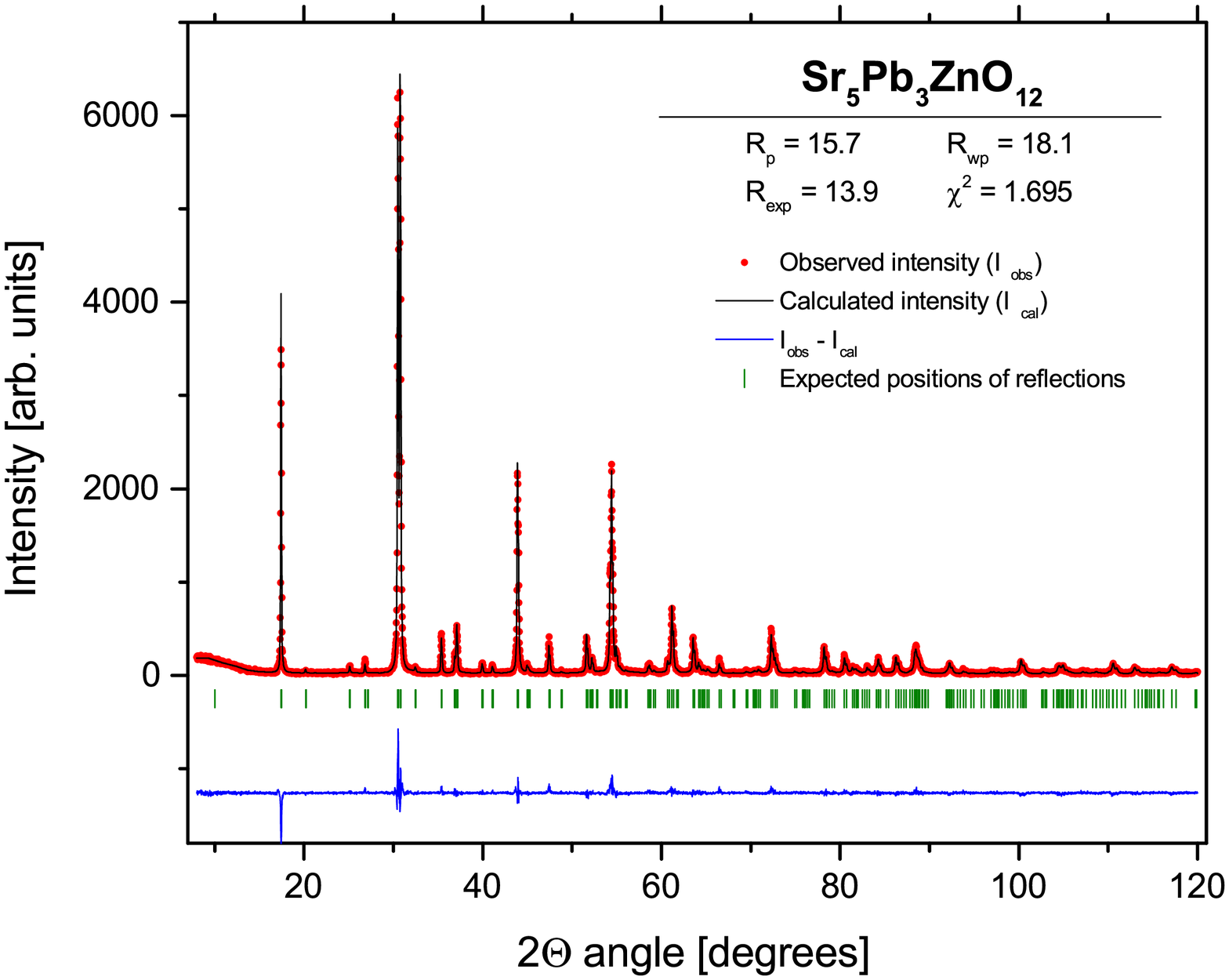}
\par\end{centering}

\begin{centering}
\label{Fig:RietveldFitPlot}
\par\end{centering}

\caption{Fit (solid blue line) of the refined structural model (Rietveld method)
to the room-temperature x-ray powder diffraction data (red circles)
for 531-Zn. Upper part \textendash{} circles, observed data, solid
line calculated intensities. The lower part (blue line) shows on the
same scale the differences between the observed and calculated pattern.
The green bars correspond to 531-Zn.}
\end{figure}

The observed x-ray spectra for Sr$_{5}$Pb$_{3}$ZnO$_{12}$, the
calculated powder-diffraction pattern, the difference between the
calculated model and experimental data, and the positions of expected
Bragg peaks are presented in Fig. \ref{Fig:RietveldFitPlot}. The
model used for refinement was derived from the structure of 531-Co\citep{yamaura_synthesis_2002}.
Cell and atomic parameters are shown in table \ref{tab:Structure}.

\begin{table}
\begin{centering}
\begin{tabular}{cccccc}
\hline 
\multicolumn{6}{c}{\textbf{Basic crystallographic data}}\tabularnewline
\hline 
\multicolumn{3}{c}{Compound formula} & \multicolumn{3}{c}{Sr$_{5}$Pb$_{3}$ZnO$_{12}$}\tabularnewline
\multicolumn{3}{c}{Molar mass {[}g/mol{]}} & \multicolumn{3}{c}{1317.0828}\tabularnewline
\multicolumn{3}{c}{Space group} & \multicolumn{3}{c}{$P\bar{6}2m$, no. 189}\tabularnewline
\multicolumn{3}{c}{a {[}\textbf{$\textrm{\AA}$}{]}} & \multicolumn{3}{c}{10.1277(1)}\tabularnewline
\multicolumn{3}{c}{c {[}\textbf{$\textrm{\AA}$}{]}} & \multicolumn{3}{c}{3.53488(4)}\tabularnewline
\multicolumn{3}{c}{Cell volume {[}\textbf{$\textrm{\AA}{}^{3}$}{]}} & \multicolumn{3}{c}{313.997(6)}\tabularnewline
\multicolumn{3}{c}{Calculated density {[}$\nicefrac{g}{cm^{3}}${]}} & \multicolumn{3}{c}{6.965}\tabularnewline
\hline 
\multicolumn{6}{c}{}\tabularnewline
\hline 
\multicolumn{6}{c}{\textbf{Atomic parameters}}\tabularnewline
\hline 
\textbf{Atom} & \textbf{x} & \textbf{y} & \textbf{z} & \textbf{B {[}$\textrm{\AA}{}^{2}${]}} & \textbf{SOF}\tabularnewline
\hline 
Sr(1) & $\frac{1}{3}$ & $\frac{2}{3}$ & $\frac{1}{2}$ & 1.71(18) & 1\tabularnewline
Sr(2) & 0.6965(4) & $0$ & $\frac{1}{2}$ & 0.83(11) & 1\tabularnewline
Pb & 0.3377(2) & $0$ & $0$ & 1.09(2) & 1\tabularnewline
Zn & $0$ & $0$ & 0.2751(51) & 6.69(55) & 0.5\tabularnewline
O(1) & 0.1779 & $0$ & $\frac{1}{2}$ & - & 1\tabularnewline
O(2) & 0.4617 & 0 & $\frac{1}{2}$ & - & 1\tabularnewline
O(3) & 0.2380 & 0.4428 & $0$ & - & 1\tabularnewline
\end{tabular}
\par\end{centering}

\caption{Crystallographic data for 531-Zn obtained using Rietveld refinement.
Figures of merit: profile residual $R_{p}=10.3\%$, weighted profile
residual $R_{wp}=13.7\%$ , expected profile residual $R_{exp}=10.49\%$
, goodness of fit $\chi^{2}=1.70$ .}
\label{tab:Structure}
\end{table}

\begin{figure}
\begin{centering}
\label{Fig:Visualisation}\includegraphics[width=8cm]{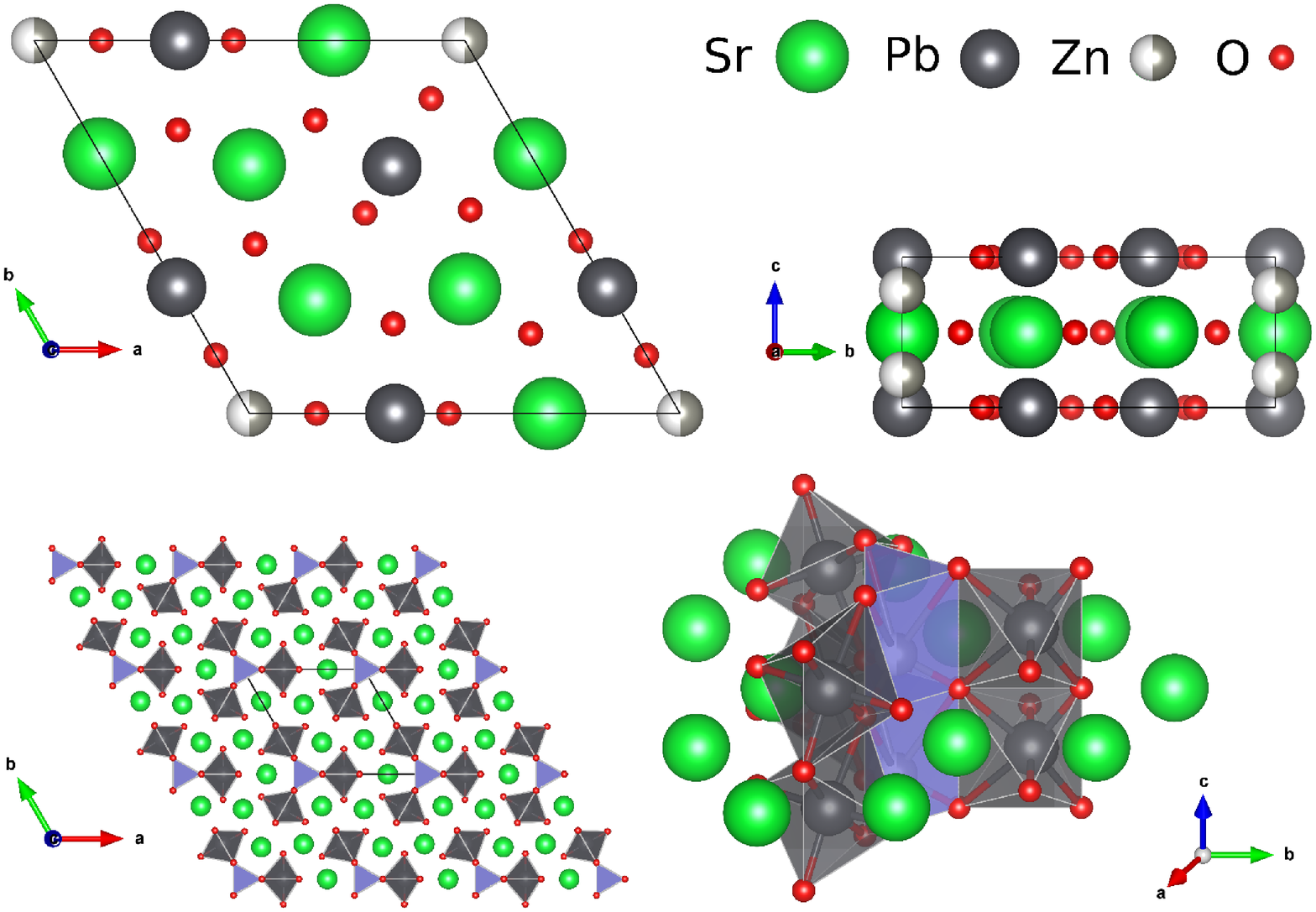}
\par\end{centering}

\caption{Visualisation of the crystal structure rendered with VESTA software\citep{momma_<i>vesta_2011}:
(a) unit cell viewed along the \textit{c} axis, (b) unit cell viewed
along the \textit{a} axis, (c) view along the \textit{c} axis revealing
the triple-symmetry of ZnO$_{6}$- PbO$_{6}$ chains surrounded by
Sr$^{2+}$, (d) a close-up of one element of the chain. Zn atoms are
half-coloured to emphasize that this position is half-occupied.}
\end{figure}

The crystal structure of the new compound was found to be similar
to other known 531-M compounds. In particular, the transition metal
ion (Zn$^{2+}$) is situated in a trigonal prismatic site, and ZnO$_{6}$
and PbO$_{6}$ polyhedra form infinite chains along the $c$-axis,
surrounded by Sr$^{2+}$ ions as is shown in Fig. \ref{Fig:Visualisation}
(c, d). The Zn position is half-occupied and the Zn-O polyhedra are
significantly distorted, as it is visible on fig \ref{Fig:Visualisation}
(d). 

The major difference between the observed and calculated intensity
is for the (300) and (111) reflections, in the 2$\Theta$ angle range
$30.5\textdegree$ and $30.8\textdegree$. This difference is possibly
caused by insufficient modeling of ZnO$_{6}$- PbO$_{6}$ chains.
The large value of thermal parameter for Zn suggests that the local
environment of Zn atoms is much more complex than it is deduced from
our refinement. Neutron powder diffraction studies of 531-Cu ran into
the same issue \citep{babu_neutron_1991}, with the authors pointing
out possible cation disorder and deviation from the Sr$_{5}$Pb$_{3}$MO$_{12}$
stoichiometry. Our EDS studies performed on single crystals, confirm
the 5:3:1 (Sr:Pb:Zn) cation ratio. Results of the iodometric titration
yield the concentration of Pb$^{4+}$ ions $x_{Pb^{4+}}=3.1(2)$ ions
per formula unit. This suggests that Pb in 531-Zn is present in $+4$
oxidation state, and because Zn can only exist in $+2$ form, the
oxygen stoichiometry is 12 atoms per formula unit. Therefore in 531-Zn
we do not expect additional oxygen atoms that partially occupy 6\textit{i}
sites as it was reported for 531-Cu \citep{yamaura_crystal_2001}
and 531-Co\citep{yamaura_synthesis_2002} compounds. The substoichiometric
concentration of the oxygen in 531-Zn is unlikely due to the fact
that syntesis was provided in air at relatively low temperature. It
is agreement with the titration results that does not suggest the
presence of Pb$^{2+}$ ions.

The physical properties of 531-Zn were first studied by the impedance
spectroscopy technique; the results are shown in Figure 4. The Nyquist
plot for the 531-Zn sample consists of one semicircle and therefore
a simple resistor-CPE model was chosen to fit the experimental data,
as shown in the Figure 4(b). The DC conductivity obtained from the
fit was $\sigma_{DC}=2.43(6)\,\cdot\,10^{-14}\:\Omega^{-1}cm^{-1}$
at $300\ K$. For the two other 531-M compounds, 531-Cu and 531-Co,
in the Nyquist plot two semicircles are visible (Fig. 4(d)) which
suggest the presence of two relaxation mechanisms. The existence of
two relaxation mechanisms as well as significantly lower resistivity
than in the case of 531-Zn has been attributed to a mixed valency
state of Cu and Co \citep{yamaura_crystal_2001,yamaura_synthesis_2002}.
The conduction mechanism in 531-Cu and 531-Co is possibly an electron
transfer between Cu$^{2+}$ and Cu$^{3+}$ ions in 531-Cu (Co$^{2+}$
and Co$^{3+}$ in 531-Co).

\begin{figure}
\begin{centering}
\includegraphics[width=10cm]{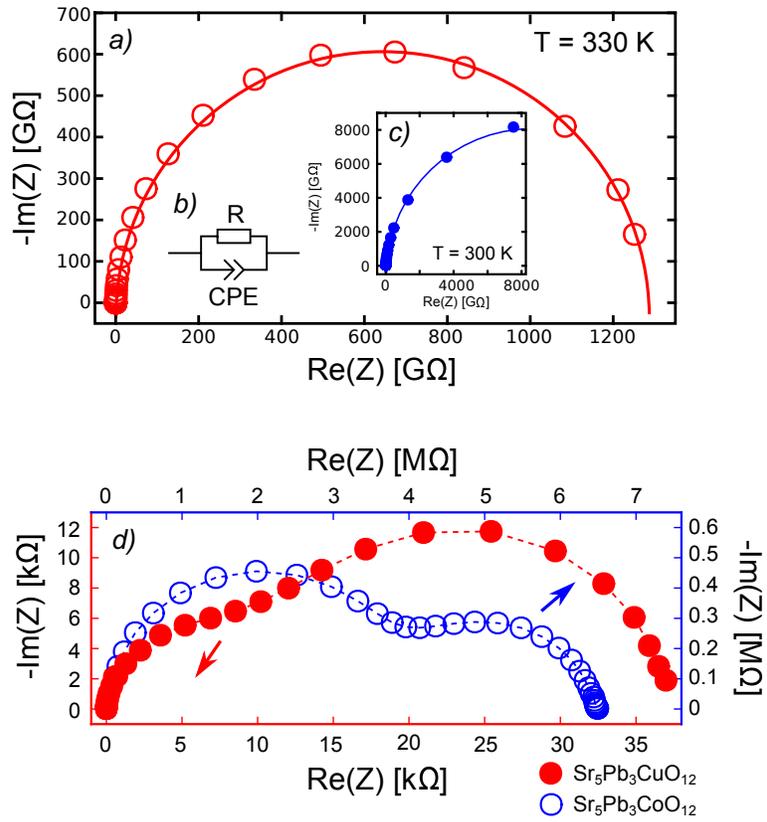}\label{fig:Nyquist}
\par\end{centering}

\caption{Nyquist plot for 531-Zn at 330 K showing one relaxation mechanism
(a), solid line is a guide for eyes. (b) the equivalent circuit used
for fitting, (c) Nyquist plot for 531-Zn at 300 K, (d) Nyquist plots
for samples of 531-Cu (red filled circles) and 531-Cu (blue circles)
synthesized and measured with the same methodology as 531-Zn.}
\end{figure}

\begin{figure}
\begin{centering}
\includegraphics[width=10cm]{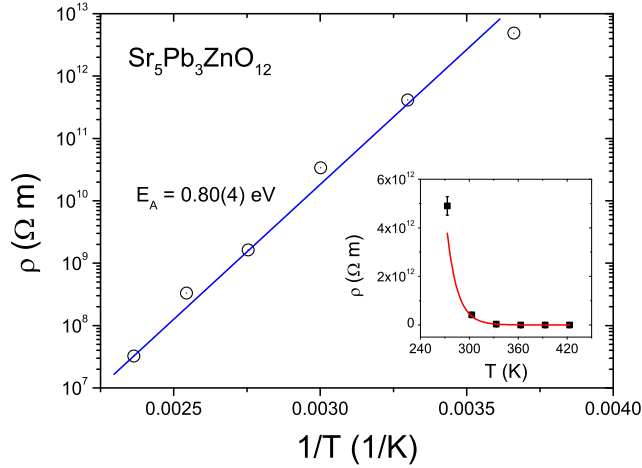}
\par\end{centering}

\caption{Natural logarithm of resistivity vs. inverse temperature (a) and resistivity
vs. temperature (b) plots. Black squares-- values obtained from Nyquist
plots, red line -- fit to the data. }
\label{fig:ActivationEnergy}

\end{figure}

DC resistivity values derived from fitted parameter were plotted against
the temperature and activation energy of the conduction process was
calculated using Arrhenius relation: 
\begin{equation}
\rho(T)=\rho_{0}\exp\left(\frac{E_{a}}{k_{B}T}\right)
\end{equation}

where: $E_{a}$ - activation energy, $k_{B}$ - Boltzmann constant.
Fig. \ref{fig:ActivationEnergy} (a) and (b) shows the plot of $\ln(\rho)=f(T^{-1})$
and $\rho=f(T)$, respectively. Activation energy of the relaxation
process has a value of $E_{A}=0.80(4)\: eV$. Relative dielectric
permittivity, $\varepsilon_{r}$, is almost frequency idependent in
a wide range of frequencies, and has a value of about $22$ at 300
K.

\begin{figure}
\begin{centering}
\includegraphics[width=9cm]{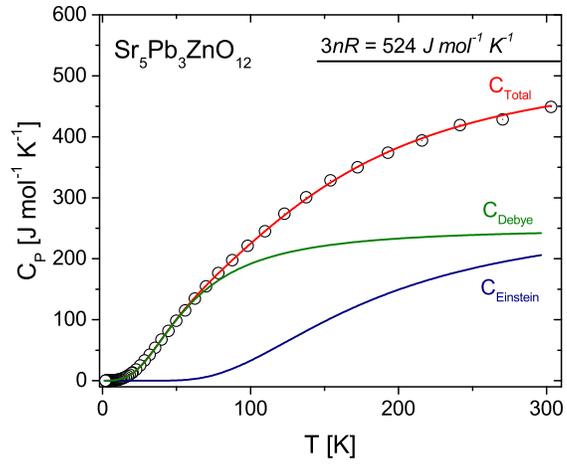}
\par\end{centering}

\caption{Specific heat vs. temperature for 531-Zn. Points -- measured specific
heat, solid red line -- fit to the measured data (eq. \ref{eq:Cp3term}),
green line -- the Debye part of total specific heat, blue line --
the Einstein part of the total specific heat. Black solid line indicates
the value calculated from the Dulong Petit Law.}
\label{fig:HeatCapacity}
\end{figure}

Fig. \ref{fig:HeatCapacity} shows the result of heat capacity measurement.

Figure \ref{fig:HeatCapacity} shows the overall temperature dependence
of the specific heat $C_{p}$. The heat capacity of the sample at
room temperature ($\sim450\: J\, mol^{-1}\, K^{-1}$) is close, but
slightly lower than the value expected from the Dulong-Petit Law ($3nR\approx524\: J\, mol^{-1}\, K^{-1}$),
where $n$ is number of atoms per formula unit and $R$ is the gas
constant). 

\begin{figure}
\begin{centering}
\includegraphics[width=9cm]{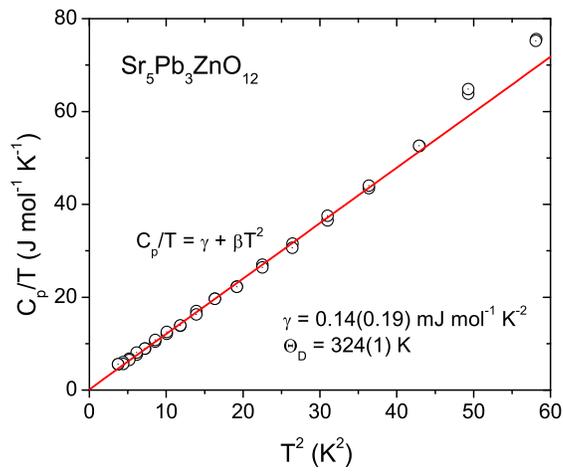}
\par\end{centering}

\caption{Specific heat divided by temperature as a function of the temperature
squared for 531-Zn. Points -- values measured in PPMS, red solid line
-- linear fit to the low temperature part of $\frac{C_{p}}{T}=f(T^{2})$. }
\label{fig:CpTvsT2}

\end{figure}

Figure \ref{fig:CpTvsT2} shows specific heat divided by temperature
as a function of temperature squared ($\frac{C_{p}}{T}=f(T^{2})$)
in the low-temperature range measured under zero magnetic field. Data
points were fit in the temperature range of $2.3\: K$ to $6.5\: K$
using the formula $\frac{C_{p}}{T}=\gamma+\beta T^{2}$. The value
of Sommerfeld coefficient obtained from the fit is $\gamma=0.14(19)\: mJ\, mol^{-1}\, K^{-2}$,
which is close to zero, as expected for a dielectric material. The
phonon specific-heat coefficient $\beta=1.194(9)\: J\, mol^{-1}\, K^{-4}$.
In a simple Debye model, at low temperature the $\beta$ coefficient
is related to the Debye temperature $\Theta_{D}$ through the relation
given in eq. \ref{eq:ThetaD}:

\begin{equation}
\Theta_{D}=\sqrt[3]{\frac{12\pi^{4}nR}{5\beta}}\label{eq:ThetaD}
\end{equation}

The Debye temperature calculated from this relation $\Theta_{D}=324(1)\: K$.
The specific heat calculated from the Debye model with this value
of $\Theta_{D}$ is not large enough to reach the experimental values
measured at the higher temperatures, therefore the data in the whole
temperature range was fit by using the formula given in eq. \ref{eq:Cp3term},
which includes the contribution of higher-energy optical modes:

\begin{equation}
C_{p}(T)=\gamma T+k\cdot C_{D}(T)+(1-k)\cdot C_{E}(T)\label{eq:Cp3term}
\end{equation}

where $C_{D}$ and $C_{E}$ are specific heat functions derived from
the Debye and Einstein models, respectively, the $k$ parameter corresponds
to the weight of the phonon contributions to the specific heat from
both models, and $\gamma$ is the Sommerfeld coefficient. Such an
approach is commonly used in case of intermetallic compounds and was
employed for the whole temperature range fitting of specific heat
data for Mg$_{10}$Ir$_{19}$B$_{16}$\citep{klimczuk_superconductivity_2006}
and UCr$_{2}$Al$_{20}$\citep{swatek_magnetic_2012}. As shown by
the low temperature fit, the electronic heat capacity, $\gamma T$,
is negligible. The temperature dependence of the specific heat for
Debye and Einstein models is given in eqs. \ref{eq:CdvsT} and \ref{eq:CevsT},
respectively:

\begin{equation}
C_{D}(T)=9nR\left(\frac{T}{\Theta_{D}}\right)^{3}\intop_{0}^{x_{D}}\frac{x^{4}\exp(x)}{(\exp(x)-1)^{2}}dx\label{eq:CdvsT}
\end{equation}

\begin{equation}
C_{E}(T)=3nR\left(\frac{\Theta_{E}}{T}\right)^{2}\exp\left(\frac{\Theta_{E}}{T}\right)\left(\exp\left(\frac{\Theta_{E}}{T}-1\right)\right)^{-2}\label{eq:CevsT}
\end{equation}

where $x=\frac{h\nu}{k_{B}T}$ and $x_{D}=\frac{\Theta_{D}}{T}$,
$\Theta_{D}$ and $\Theta_{E}$ being Debye and Einstein temperatures,
respectively. The fit, plotted as a solid red line on fig. 6, gives
$k=48\%$ of the weight to the Debye contribution with $\Theta_{D}=237(2)\: K$
and $(1-k)=52\%$ to Einstein term with $\Theta_{E}=556(4)\: K$.

\section{Conclusions}

A novel solid-state compound, Sr$_{5}$Pb$_{3}$ZnO$_{12}$ (531-Zn),
was synthesized and characterized. The structure of the new compound
closely resembles the structures of the other 531-M oxides. In contrast
to the 531-M (M=Co, Cu, Ni) compounds, the oxygen stoichiometry for
531-Zn is 12 per formula unit and no additional oxygen atoms partially
occupying the 6\textit{i} sites are expected. A single-crystal growth
experiment showed that needle-like crystals can be grown using simple
mineralization method with NaCl as the mineralizer. Process parameters
(especially cooling rate) should be optimised to grow crystals suitable
for measuring physical properties. 

531-Zn was found to be a dielectric with rather high relative permittivity
($\varepsilon_{r}=22$ at 300K) and with one relaxation mechanism
visible in the examined range of temperature and frequency. This is
in contrast with results for 531-Cu and 531-Co, where two mechanisms
are observed. The conductivity activation energy for 531-Zn was estimated
by using the Arrhenius fit and its value is $E_{A}=0.80(4)\: eV$.
As expected for insulators, the Sommerfeld coefficient ($\gamma)$
estimated from the heat capacity measurement, is equal to zero (within
the uncertainty). The low temperature fit gives the Debye temperature
$\Theta_{D}=324(1)\: K$ for 531-Zn.

Future studies would involve doping of 531-Zn to determine whether
an insulator-to-metal transition can be induced. One unconfirmed report
of a metal-insulator transition accompanied by a strong structural
modulation and possible superconductivity in a heavily La-doped 531-Cu
compound has been published\citep{che_metal-insulator_2000} but no
further studies have been conducted on this apparent superconductor
to the authors knowledge.

\section*{Acknowledgements}

The project was partially supported by the Polish National Science
Centre (Decision no. DEC-2012/07/E/ST3/00584).

The authors acknowledge helpful discussions with Robert J. Cava (Princeton
University).

\section*{References}

\bibliographystyle{elsarticle-num}
\addcontentsline{toc}{section}{\refname}\bibliography{531-all}

\end{document}